# Large-signal model of the Metal-Insulator-Graphene diode targeting RF applications

Francisco Pasadas, Mohamed Saeed, *Student Member, IEEE*, Ahmed Hamed, Zhenxing Wang, Renato Negra, Daniel Neumaier and David Jiménez

*Abstract*—We present a circuit-design compatible large-signal compact model of metal-insulator-graphene (MIG) diodes for describing its dynamic response for the first time. The model essentially consists of a voltage-dependent diode intrinsic capacitance coupled with a static voltage-dependent current source, the latter accounts for the vertical electron transport from/towards graphene, which has been modeled by means of the Dirac-thermionic electron transport theory through the insulator barrier. Importantly, the image force effect has been found to play a key role in determining the barrier height, so it has been incorporated into the model accordingly. The resulting model has been implemented in Verilog-A to be used in existing circuit simulators and benchmarked against an experimental 6-nm $TiO_2$ barrier MIG diode working as a power detector.

## I. INTRODUCTION

VERTICAL metal-insulator-graphene (MIG) diodes for radio-frequency applications have been recently demonstrated [1], [2]. The key property exploited in this device is the work function tunability of graphene, resulting in a large modulation of the insulator barrier height. This kind of diodes shows excellent high on-current density, high asymmetry, strong maximum nonlinearity and large maximum responsivity, outperforming state-of-the-art metal-insulator-metal (MIM) diodes [1]. MIG diodes have been also demonstrated to perform well as key devices of different circuits such as power detectors, mixers, and six-port receivers [1], [3]–[7]. To simulate circuits based on the diode nonlinearity, small-signal models, as the ones reported in [4], [7], are no longer suitable, so a large-signal is required instead. The goal of this work is to develop and implement such a large-signal model in Verilog-A to provide a TCAD tool for circuit design. Our model is physics-based, contrary to the one reported in [8], which is semiempirical.

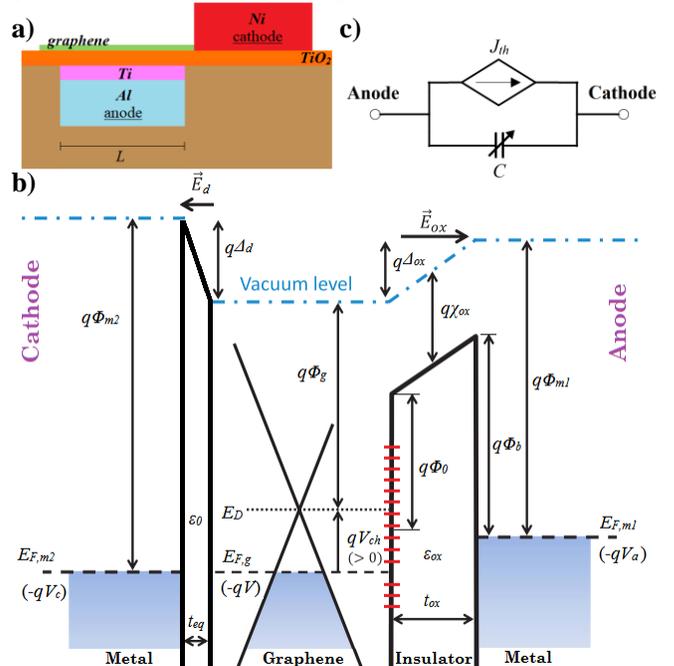

Fig. 1 a) Cross-section of a MIG diode. b) Band diagram of a metal-insulator-graphene-metal structure. $q$ is the elementary charge; $\Phi_{m1}$, $\Phi_{m2}$ and $\Phi_g$ are the anode metal, cathode metal and graphene work functions, respectively; $\chi_{ox}$ is the electron affinity of the insulator; $E_{F,m1} = -qV_a$, $E_{F,m2} = -qV_c$, and $E_{F,g}$ are the Fermi energy levels in both metals and graphene, respectively; $E_D$ is the Dirac energy where the conduction and valence bands of graphene touch each other; $V_{ch} = (E_D-E_{F,g})/q$ is the graphene chemical potential; $V_c$ and $V_a$ are the applied cathode and anode voltages, respectively; $E_{ox}$ and $E_d$ are the electric fields in the oxide and dipole layer, respectively; $\varepsilon_{ox}$ and $\varepsilon_0$ are the oxide and vacuum permittivities, respectively; and $t_{ox}$ and $t_{eq}$ are the oxide thickness and equilibrium dipole separation distance. c) Intrinsic large-signal MIG diode equivalent circuit formed by a voltage-dependent current source and a voltage-dependent intrinsic capacitance.

## II. COMPACT LARGE-SIGNAL MODEL OF THE MIG DIODE

### A. *Electrostatics of the MIG diode*

The band diagram of the metal/insulator/graphene/metal heterostructure, corresponding to the device sketched in Fig. 1a, has been shown in Fig. 1b. Because of the charge transfer between the cathode metal and graphene, a dipole layer of size $t_{eq}$ inside the equilibrium separation distance is set up [9]. The cathode metal and graphene are in equilibrium and, hence, the Fermi energies are aligned. Considering the energy potential loops at both graphene interfaces, together with the charge

Manuscript received Month Day, Year; revised Month Day, Year; accepted Month Day, Year. This work has received funding from the European Union's Horizon 2020 Research and Innovation Programme under grant agreement No 785219 GrapheneCore2 and from MINECO under Grant TEC2015-67462-C2-1-R.
(*Corresponding author: Francisco Pasadas.*)
F. Pasadas and D. Jiménez are with the Departament d'Enginyeria Electrònica, Escola d'Enginyeria, Universitat Autònoma de Barcelona, 08193 Bellaterra, Barcelona, Spain (email: francisco.pasadas@uab.es).
M. Saeed, A. Hamed, and R. Negra are with High Frequency Electronics, RWTH Aachen University, 52062 Aachen, Germany.
Z. Wang, and D. Neumaier are with AMO GmbH, 52074 Aachen, Germany.
Color versions of one or more of the figures in this paper are available online at http://ieeexplore.ieee.org.
Digital Object Identifier:

conservation law, results in [10]:

$$\begin{aligned} \phi_{m1} - (\phi_g + V_{ch}) &= V_a - V_c - \Delta_{ox} \\ \phi_{m2} - (\phi_g + V_{ch}) &= \Delta_d \\ Q_{m1} + Q_{m2} + Q_g &= 0 \end{aligned} \quad (1)$$

In (1), the term $\Delta_d$ is the potential drop across the dipole layer which can be expressed as $\Delta_d = \Delta_{tr} + \Delta_{ch}$, where $\Delta_{tr}$ corresponds to the charge transfer and $\Delta_{ch}$ to the chemical potential interaction. The latter describes the short range interaction from the overlap of the metal and graphene wave functions [9], [11]. $Q_{m1} = C_{ox}\Delta_{ox} + Q_f$ gives the charge per unit area induced in the anode metal, where the charge density, $Q_f$, represents any fixed charges within the oxide layer [12]; $Q_{m2} = -C_d\Delta_{tr}$ gives the charge per unit area induced in the cathode metal. The capacitive coupling to the cathode metal per unit area is $C_d = \varepsilon_0/t_{eq}$ and the geometric oxide capacitance per unit area is $C_{ox} = \varepsilon_{ox}/t_{ox}$. The graphene charge per unit area can be written as $Q_g = Q_{net}[V_{ch}] + Q_{it}[V_{ch}] + Q_0$, where $Q_0$ is the charge density due to possible chemical doping [13]; $Q_{it} = C_{it}V_{ch} + Q'$ represents the charge density due to the possible presence of interface traps, where $C_{it}$ is the interface-trap capacitance per unit area and $Q' = C_{it}(\Phi_g-\chi_{ox}-\Phi_0)$, where $\Phi_0$ is the charge neutrality level; and $Q_{net}$ is the overall net mobile sheet charge density in the channel:

$$Q_{net} = q(p-n) = \frac{2q(k_BT)^2}{\pi(\hbar v_F)^2}\left(F_1\left[\frac{qV_{ch}}{k_BT}\right] - F_1\left[-\frac{qV_{ch}}{k_BT}\right]\right) \quad (2)$$

Here $p$ and $n$ are the hole and electron carrier densities, respectively; $F_1$ is the first-order Fermi-Dirac integral and has been approximated using elementary mathematical functions [14]–[16] because no closed-form solution exists.; $k_B$ is the Boltzmann constant; $T$ is the temperature; $\hbar$ is the reduced Planck's constant; and $v_F \approx 10^6$ m/s is the Fermi velocity.

Combining the equations in (1), the electrostatics of the heterostructure can be described by the following equation:

$$C_{ox}(V_a - V_c) + (C_{ox} + C_d + C_{it})V_{ch} + Q_{net}[V_{ch}] = Q^* \quad (3)$$

where $Q^* = -(Q' + Q_0 + Q_f + C_{ox}(\Phi_g - \Phi_{m1}) + C_d(\Delta_{ch} + \Phi_g - \Phi_{m2}))$.

*B. Vertical electron transport through the MIG diode*

While the in-plane electron transport properties of graphene and other 2D materials have been extensively studied, the out-of-plane transport, such as electron vertical emission from the plane has remained relatively less explored [17] and only recently has received attention due to the emergence of various 2D-material-based van der Waals heterostructures [18], [19]. Due to the unusual quasiparticle dynamics in graphene, the traditional emission equations for the conduction mechanisms are no longer valid. Aside from the barrier height at the electrode/insulator interface, the effective mass of the conduction carriers is also a key factor, thus being critical to take into account the linear energy dispersion of graphene in the vertical emission description [20]–[23].

The usual temperature-dependent static current flow through a MIG diode can be attributed to thermionic emission of carriers above the barrier. It can be described by the modified Dirac-Schottky model that accounts for the different density-of-states of graphene (2D) and metal (3D) [22]:

$$J_{th} = \frac{qk_B^3T^3}{\pi\hbar^3 v_F^2} e^{-\frac{q\phi_b - \frac{\delta_p^2}{2k_BT}}{k_BT}} \left(e^{\frac{q(V_a-V_c)}{\eta k_BT}} - 1\right) \quad (4)$$

where $\Phi_b$ is the Schottky barrier height (SBH) (see Fig. 1b); $\delta_p$ estimates the spreading of the Fermi level, and $\eta$ is the ideality factor. Note that a scaling of $T^3$ is predicted, different from the Richardson-Dushman scaling of $T^2$ for bulk materials.

The image force (IF) effect has been found to be a crucial ingredient to explain the experimental results, strongly impacting on the SBH. A good approximation for the SBH that includes the image potential along the insulator thickness ($y$-coordinate) is given by [24], [25]:

$$\phi[y] = \phi_b - \Delta_{ox}\frac{y}{t_{ox}} - 1.15\frac{\ln[2]}{8\pi}\frac{q}{C_{ox}}\frac{1}{y(y-t_{ox})} = \phi_b - \text{SBL} \quad (5)$$

The highest $\phi$ that defines the insulator barrier height is found requiring $d\phi[y^*]/dy = 0$, where $\phi[y^*]$ replaces $\phi_b$ in (4) to account for the Schottky barrier lowering (SBL) produced by the IF effect.

*C. Charge-based intrinsic capacitance of the MIG diode*

An accurate modeling of the intrinsic capacitances of a device requires an analysis of the charge distribution in the channel versus the terminal voltages. The charges $Q_a$ and $Q_c$ associated with the anode and cathode, respectively, are defined as $Q_a = WLQ_{net}$. $Q_c = -Q_a$ guarantees charge conservation. Thus, a two-terminal device can be modeled with four intrinsic capacitances, where $C_{ij}$ describes the dependence of the charge at terminal $i$ with respect to a varying voltage applied to terminal $j$ assuming that the voltage at the other terminal remains constant.

$$C_{ij} = -\frac{\partial Q_i}{\partial V_j} \quad i \neq j \qquad C_{ij} = \frac{\partial Q_i}{\partial V_j} \quad i = j \quad (6)$$

where $i$ and $j$ stand for $a$ and $c$. However, charge-conservation and a reference-independent model results in only one independent capacitance for describing the two-terminal diode:

$$C = \frac{C_{ox}C_q}{C_{ox} + C_q + C_d + C_{it}} \quad (7)$$

where $C_q = dQ_{net}/dV_{ch}$ is the quantum capacitance of graphene, calculated by the following analytic expression [14]:

$$C_q = \frac{2q^2 k_B T}{\pi(\hbar v_F^2)} \ln\left[2\left(1 + \cosh\left[\frac{qV_{ch}}{k_BT}\right]\right)\right] \quad (8)$$

The resulting intrinsic large-signal model of the MIG diode, shown in Fig. 1c, has been implemented in Verilog-A and included in the circuit simulator Keysight© ADS.

III. MODEL ASSESSMENT

In this section, we first compare the DC characteristic of an experimental MIG diode with our model outcome. Next, an RF power detector based on such device is simulated and later compared with measurements [3]. The experimental MIG diode (graphene length $L$ = 2 μm and width $W$ = 80 μm) consists of 6-nm TiO$_2$ embedded between an Al/Ti metal electrode [1], acting as the anode, and Ni electrode acting as the cathode [26]. The cross-section of the experimental device is shown in Fig. 1a. The fabrication process is reported elsewhere [1]. The parameters used for describing the device in [3] according to the model presented in Section II are given in Table I. In addition to the intrinsic device, the appropriate extrinsic network must be included to describe the operation of the device. Fig. 2 shows the equivalent circuit of the device, where $C_{fringe}$ ($C_{fp}$) represents the parasitic diode (pad) fringe capacitance, $L_{a,ext}$ and $R_{a,ext}$ ($L_{c,ext}$ and $R_{c,ext}$) are the anode (cathode) extrinsic inductances and resistances due to the pads, respectively; and $C_{pa}$ ($C_{pc}$) represents the anode (cathode) parasitic capacitance due to the substrate. The latter elements have been obtained from a de-embedding procedure. On the other hand, the cathode metal – graphene contact resistance, $R_{contact}$, has been extracted from the fitting of the experimental results and been estimated to be of 2.0 kΩ·µm.

TABLE I. INPUT PARAMETERS OF THE MIG DIODE UNDER TEST.

| | | | |
|---|---|---|---|
| $L$ | 2 µm | $\Phi_g$ | 4.5 eV |
| $W$ | 80 µm | $\Phi_{m,Ti}$ | 4.33 eV |
| $C_{ox}$ | 7.1 mF/m$^2$ | $\Phi_{m,Ni}$ | 5.05 eV |
| $C_d$ | 43.2 mF/m$^2$ | $\chi_{ox}$ | 3.66 eV |
| $C_{it}$ | 0 F/m$^2$ | $\delta_p$ | 0 eV |
| $Q^*$ | -31.4 mC/m$^2$ | $\eta$ | 1 |

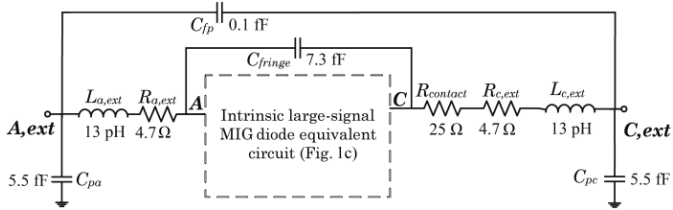

Fig. 2 Topology of the equivalent circuit of the MIG diode under test including extrinsic elements extracted after applying a de-embedding procedure by using open structures. The intrinsic part corresponds to the equivalent circuit depicted in Fig. 1c.

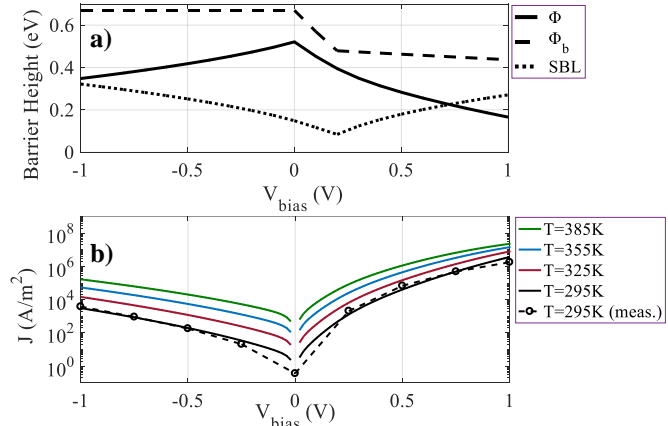

Fig. 3 a) Barrier height for electron transport ($\Phi$, $\Phi_b$ and SBL) at 295K. b) Current density-voltage ($J$-$V_{bias}$) simulations for different temperatures (solid lines) and measurements at 295K (circles) of the device described in Table I.

Fig. 3 shows a comparison between the simulated results and the measured current density-voltage ($J$-$V_{bias}$), where $J=J_{th}$ and $V_{bias}=V_{A,ext}-V_{C,ext}$. The inset of Fig. 3 shows the barrier height for electron transport before ($\Phi_b$) and after the IF correction ($\Phi$) together with the difference between those magnitudes (SBL). When the diode is in reverse bias, the current dramatically deviates from the constant behavior that would be predicted by $\Phi_b$. In direct bias, the IF effect also plays a crucial role in scaling up the total current flow and cannot be disregarded.

Now the MIG diode is used as the active block of a power detector. Diodes convert high frequency energy to DC by way of their rectification properties, which arise from their nonlinear $J$-$V_{bias}$ characteristics. Fig. 4 shows the schematics of the RF power detector proposed in [3]. The incident RF input power, $P_{in}$, is coupled with a zero DC bias to the diode using a bias tee. Then, an impedance matching (IM) stage is necessary to guarantee that optimum RF power is delivered to the diode ($S_{11}$ < -10 dB). At the diode output terminal, a low-pass filter (LPF) is attached with a cutoff frequency of 160 MHz, which is used to attenuate the AC signal. A DC output voltage is measured across a high-resistance load (10 MΩ representing the input impedance of a voltmeter) attached to the LPF output. The bias tee, the IM, and the LPF have been simulated by using equivalent circuits of the lumped elements.

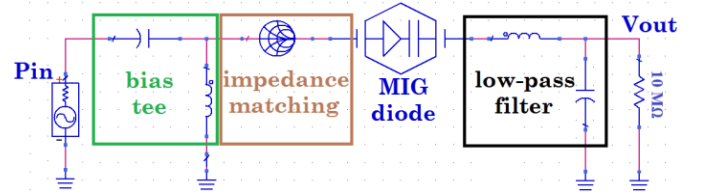

Fig. 4 Schematics of a MIG diode-based power detector. The "MIG diode" symbol contains the network depicted in Fig. 2.

The measured and simulated DC output signal versus input RF power at 2.45 GHz have been plotted in Fig. 5, showing good agreement. The experimental results show a linear dynamic range of 50 dB. The simulated (measured) DC and RF responsivity of the detector are 24.7 (26) V$^{-1}$ and 43.6 (42) V/W, respectively. For input powers lower than -40 dBm, the experimental output voltage reaches a minimum value ~ 1 µV probably set by a noise power comparable with the signal power. For input powers higher than 5 dBm, the model predicts the end of the square-law detection.

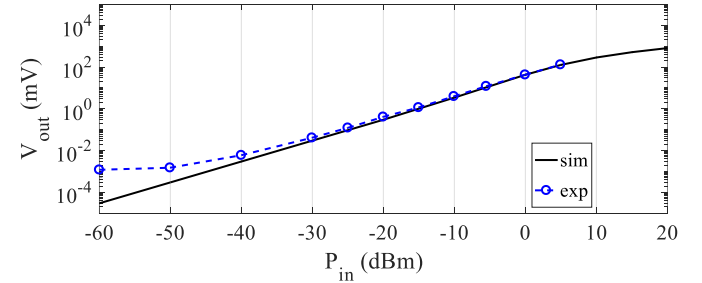

Fig. 5 DC output voltage versus RF input power at room temperature. The corresponding high frequency power detection responsivity can be extracted from the slope of the linear relationship.

IV. CONCLUSIONS

In conclusion, we have presented a physics-based large-signal compact model of the MIG diode implemented in Verilog-A. It can predict the bias dependence of the barrier height and correctly describes the nonlinear behavior of MIG diodes enabling the simulation of high-frequency large-signal operation of complex circuits based on such devices.